\documentstyle[prl,aps,preprint,psfig]{revtex}

\newcommand{\tauh}{\tau_{\scriptscriptstyle H}}

\newcommand{\geqa}{\stackrel{>}{\scriptstyle \sim}}
\newcommand{\kf}{k_{\scriptscriptstyle F}}

\newcommand{\ef}{E_{\scriptscriptstyle F}}
\newcommand{\ub}{\overline{U}}
\newcommand{\ut}{\widetilde{U}}
\newcommand{\utr}{\widetilde{U}_{reg}}
\newcommand{\utc}{\widetilde{U}_{ch}}

\newlength{\figwidth}
\title{Nuclear masses: evidence of order--chaos coexistence}
\author{O. Bohigas and P. Leboeuf}
\address{Laboratoire de Physique Th\'eorique et Mod\`eles Statistiques,
B\^at. 100, \\ Universit\'e de Paris-Sud, 91405 Orsay Cedex, France}

\begin{document}

\maketitle {\begin{abstract} Shell corrections are important in the
determination of nuclear ground--state masses and shapes. Although general
arguments favor a regular single--particle dynamics, symmetry--breaking and
the presence of chaotic layers cannot be excluded. The latter provide a
natural framework that explains the observed differences between experimental
and computed masses.
\end{abstract}}
\vspace{3cm}
\hspace{1.1cm} PACS numbers: 21.10.Dr, 24.60.Lz, 05.45.Mt
\narrowtext

\pagebreak
\tightenlines

Different approaches have been developed to reproduce the systematics of the
observed nuclear masses. Some of them are of microscopic origin, while others,
more phenomenological, are inspired from liquid drop models or Thomas--Fermi
approximations. Often, the total energy is expressed as the sum of two terms
\begin{equation} \label{u}
U (Z, N, x) = \ub (Z, N, x) + \ut (Z, N, x) \ ,
\end{equation}
where $Z$ and $N$ are the proton and neutron numbers, respectively, and $x$
represents a set of parameters that define the shape of the atomic nucleus.
The first term $\ub$ describes the bulk (macroscopic) properties of the
nucleus, and contains all the contributions that vary smoothly with proton and
neutron numbers. Its typical value is close to 8 MeV/$A$ throughout the
periodic table, where $A = N + Z$ is the mass number. The second term $\ut$
describes shell effects, related to the shape--dependent microscopic
fluctuations of the nuclear wavefunctions \cite{bm}. A method to incorporate 
these effects in a consistent manner was originally proposed by Strutinsky 
\cite{st1}.

Global nuclear mass calculations have been pursued over the years with
increasing precision \cite{mnms,apdt}. Despite the numerous parameters
contained in the models, the accuracy of the results obtained and the quality
of the predictions are impressive. Different models and potentials yield
similar results, and give an accuracy of $5 \times 10^{-4}$ for a
medium--heavy nucleus whose total (binding) energy is of the order of 1000
MeV.

One may ask whether the difference between measured and computed masses has a
particular significance, since it seems to be quite model independent. Among
other possibilities, a natural explanation would come from many-body effects
not included in the mean--field scheme. Our purpose here is to show that there
is a very natural and appealing dynamical explanation. We will argue that
$\ut$ may be splitted into two parts, $\ut = \utr + \utc$, and that the
present calculated masses are able to correctly reproduce only $\utr$. The two
contributions $\utr$ and $\utc$ originate in regular and chaotic components of
the motion of the nucleons, respectively. Although our calculations are done
within a mean--field approximation, the final result for the fluctuations
produced by the chaotic part of the motion are in fact of a much more
general validity, and may be interpreted as arising from the residual
interactions.

With essentially no free parameters, this dynamical symmetry--breaking
mechanism provides a quantitative description of the basic observations on
nuclear masses. First, the amplitude of $\utc$ is much smaller than that of
$\utr$. Second, the typical size of the fluctuations $\utr$ have a constant
amplitude as a function of the number of nucleons. The RMS of $\utr$ is found
to be of about 3 MeV, which is in good agreement with the typical size of the
difference $\delta U = U_{exp} - \ub$ between the experimental and the
computed bulk properties of the nuclei \cite{mnms}. Third, the RMS of
$\utc$ is of order $0.5$ MeV, in good agreement with the typical size of the
difference $\delta U = U_{exp} - U_{calc}$ between the experimental and the
calculated masses including shell effects \cite{mnms,apdt}. The mass-number
dependence of this difference is also well reproduced.

From a semiclassical point of view, shell effects are interpreted as
modulations in the single--particle spectrum produced by the
periodic orbits of the corresponding classical dynamics \cite{bb,sm}.
Given the single--particle energy levels $E_j (x)$ computed from some
Hamiltonian $H (x)$, the level density $\rho (E,x) = \sum_j \delta \left[ E -
E_j(x) \right]$ is approximated by
\begin{equation} \label{rho}
{\rho} (E,x) = \overline{{\rho}} (E,x) + \widetilde{{\rho}} (E,x) \ .
\end{equation}
The quantity $\overline{{\rho}}$ is the average density of states, 
whereas the oscillating part is expressed as
\begin{equation} \label{rhot}
\widetilde{{\rho}} = 2 \sum_p \sum_{r=1}^{\infty} A_{p,r} (E,x) \cos
\left[ r S_p(E,x)/\hbar+\nu_{p,r} \right] \ . 
\end{equation}
The sum is over all the primitive periodic orbits $p$ (and their repetitions
$r$) of the single--particle Hamiltonian. Each orbit is characterized by 
its action $S_p$, stability amplitude $A_{p,r}$, and Maslov index 
$\nu_{p,r}$. 

The shell correction to the nuclear mass is computed by inserting the
oscillatory part of the density of states into the expression of the energy,
$\ut (x,A,T)= \int dE~E~\widetilde{{\rho}} (E,x)~f(E,\mu,T)$, with $f$ the
Fermi function. In a semiclassical expansion, the leading order of the
integral comes from the energy dependence of the action, and the
dependence of the prefactors can be ignored. Setting moreover the 
temperature $T$ to zero, one obtains
\begin{equation} \label{ut}
\ut (x,A) = 2 \hbar^2 \sum_p \sum_{r=1}^{\infty}
\frac{A_{p,r}}{r^2 ~ \tau_p^2} \cos(r S_p/\hbar+\nu_{p,r})  \ ,
\end{equation}
where $\tau_p=\partial S_p/\partial E$ are the periods of the periodic orbits.
The classical functions entering this expression {\it are evaluated at the
Fermi energy} $\ef$, related to the mass number and shape parameters through
the condition $\int_0^{\ef} \overline{\rho} (E,x) ~ dE = A$. According to
Eq.~(\ref{ut}), each periodic orbit produces a modulation or bunching of the
single--particle states on an energy scale $h/\tau_p$. The presence of
fluctuations in the total energy is therefore a very general phenomenon that
occurs for an arbitrary Hamiltonian, irrespective of the nature of the
corresponding classical dynamics. However, their importance (i.e., their
amplitude) depends strongly on the properties of the dynamics, and in
particular on its chaotic or regular character \cite{sm,lm3}. In order to see
this explicitly, we compute the variance of the fluctuations of the energy.
The average is performed over a mass number window around a given nucleid, as
when analyzing experimental data. The size of the window is taken to be small
with respect to macroscopic quantities like the Fermi energy, but large
compared to the typical oscillatory scale of $\ut$. This average is denoted by
brackets. From Eq.~(\ref{ut}) we obtain for the variance a double sum over
periodic orbits labeled by the indices $p$ and $p'$. By considering the
dependence of the density of periodic orbits with their period, it has been
shown \cite{lm3,lmb} that the dominant contribution to the double sum is given
by the diagonal terms $p=p'$, $r=r'$. The resulting sum is convergent. Using
moreover the semiclassical definition of the form factor,
$$
K (\tau) = 2 h \int_0^\infty d\epsilon \cos (\epsilon \tau/\hbar) 
\langle \widetilde{\rho} (E-\epsilon/2) \widetilde{\rho} (E+\epsilon/2)
\rangle
$$
whose diagonal part is
\begin{equation} \label{ffd}
K_D (\tau)= h^2  \sum_{p,r} A_{p,r}^2 ~ \delta 
\left( \tau - r \tau_p \right) \ ,
\end{equation}
the energy variance may be written as
\begin{equation} \label{ut2f}
\langle \ut^2 \rangle \approx \frac{\hbar^2}{2 \pi^2}
\int_0^{\infty} \frac{d\tau}{\tau^4} ~ K_D (\tau) \ .
\end{equation}
This is the basic equation that provides the starting point for our analysis.
An accurate evaluation of $\langle \ut^2 \rangle$ requires the knowledge of
the periodic orbits, and therefore of the Hamiltonian. However, we do not want
to make a calculation for a specific system. Instead we are interested in
making general statements based on the statistical behavior of the
single--particle orbits. That behavior depends on the regular or chaotic
nature of the dynamics.

General considerations suggest that the single--particle motion in the nucleus
should be dominated by regular orbits. This is based on the following
arguments. There are several factors that are important when considering the
amplitude of the shell effects as given by Eq.~(\ref{ut}). The first one is
the prefactor $A_{p,r}/\tau_p^2$. It implies that the more stable and short
the periodic orbit is, the larger its contribution. The second is associated
with the interference properties. Shells are wave effects: they are determined
by a superposition of oscillatory contributions from different periodic
orbits. The third factor, closely related to the previous one, is the phase
space structure of the periodic orbits, i.e., their degeneracy. It depends on
the existence of conserved quantities. In fully chaotic dynamics (where only
the energy is conserved), the periodic orbits are isolated and unstable. In
contrast, if there are as many conserved quantities as degrees of freedom, the
motion is regular (or integrable) and periodic orbits come in degenerate
families, all orbits having the same properties (action, stability, etc)
within a family. Since many orbits contribute coherently, this degeneracy
leads to an enhancement of the size of the shell corrections. In its
equilibrium state, and for a given number of nucleons, the nucleus will adapt
its shape in order to minimize its energy. This minimization procedure, that
takes also into account the contributions from $\ub$, favors regular
single--particle dynamics that produce strong effects.

Equation (\ref{ut2f}) allows to estimate the typical fluctuations arising from
a regular dynamics. In the range $\tau_{min} \ll \tau \ll \tauh$, where
$\tau_{min}$ is the {\it period of the shortest periodic orbit} in the
semiclassical sum (\ref{ut}) and $\tauh = h \overline{\rho}$ is the {\it
Heisenberg time}, the form factor of the regular single--particle levels is
given by $K_D (\tau) = \tauh$ \cite{berry}. This approximation is not valid
for short times, of the order of $\tau_{min}$. The reason is that for $\tau
\approx \tau_{min}$, Eq.~(\ref{ffd}) produces system specific delta function
peaks. If for simplicity we ignore this and extrapolate the statistical
behavior down to $\tau \approx \tau_{min}$, and moreover use $K_D (\tau) = 0$
for $\tau < \tau_{min}$ (cf. Eq.~(\ref{ffd})), then Eq.~(\ref{ut2f}) leads to
\begin{equation}\label{ut2r}
\langle \utr^2 \rangle = \frac{1}{24 \pi^4} \ g \ E_c^2 \ . 
\end{equation}
The quantity $E_c$ is the {\it energy associated with the shortest
periodic orbit},
\begin{equation}\label{ec}
E_c = h/\tau_{min} \ ,
\end{equation}
whereas the parameter $g$ measures this energy in units of the
{\it single--particle mean level spacing} $\delta = 1/\overline{\rho}$
\begin{equation}\label{g}
g = E_c/\delta =\tauh/\tau_{min} \ ;
\end{equation}
$g$ counts the number of single--particle states on the scale $E_c$. To
compute $E_c$ we need the period of the shortest orbit. Its length will
typically be two or three characteristic nuclear dimensions. For simplicity we
assume it to be two. Then, for a flat mean--field, like the Woods--Saxon
potential, we have $E_c = \pi \ef /\kf r$, where $r$ is the nuclear radius and
$\kf$ is the Fermi wave vector. Since $r \approx 1.1 A^{1/3}$ fm, we arrive at
\begin{equation}\label{ecn}
E_c = \frac{77.5}{A^{1/3}} \ {\rm MeV} .
\end{equation}
$E_c$ is the largest oscillatory energy scale where coherent bunching effects
occur in any thermodynamic quantity. This result should be contrasted to the
traditional shell--effect estimate based on the harmonic oscillator potential,
$E_c = \hbar w_0 \approx 40/A^{1/3}$ MeV \cite{bm}.

Since $\delta \approx 2 \ef /3 A = 25/A$ MeV, Eq.~(\ref{g}) leads to
\begin{equation}\label{gn}
g= \pi A^{2/3} .
\end{equation}
From these results we obtain for the typical fluctuations $\sigma_{reg} =
\sqrt{\langle \utr^2 \rangle}$ of a nucleus with a regular single--particle
dynamics the value
\begin{equation}\label{sigr}
\sigma_{reg} = 2.84 \ {\rm MeV} \ ,
\end{equation}
an expression independent of the number of nucleons. This estimate is in good
agreement with the $\sim 3$ MeV observed when $\ub$ is subtracted out from the
experimental or calculated values. The use of the real nuclear
mean level spacing $\delta$ to compute (\ref{gn}) (i.e., including spin and
isospin degrees of freedom) amounts to treating the different contributions as
uncorrelated, and therefore provides a lower bound.

In deriving Eq.~(\ref{sigr}) a regular single--particle motion of the nucleons
has been assumed. Even though we have stressed that a regular motion favors the
minimization of the total energy, in the full many--body problem
nothing guarantees the perfect integrability of the semiclassical
dynamics. If some dynamical symmetries are broken, chaotic components will
coexist with regular single--particle motion. The consequences of their 
presence on the behavior of the nuclear masses will now be discussed.

From a semiclassical point of view, the most simple approximation that can be
made in the generic case of a mixed dynamics, when regular orbits coexist with
chaotic layers, is to split the sum over periodic orbits in Eq.~(\ref{ut})
into two terms, one from the regular part, another from the chaotic part. The
shell energy is now written as
\begin{equation} \label{uts}
\ut = \utr + \utc .
\end{equation}
The two terms are, from a statistical point of view, independent, $\langle
\utr \utc \rangle =0$. This happens because, as already pointed out, the
dominant contribution to the energy comes from the short orbits, with $\tau_p
\ll \tauh$. Since the orbits contributing to each term are different, the
cross products vanish by the averaging procedure (assuming that the actions of
the orbits are incommensurable). The decomposition (\ref{uts}) identifies the
difference between measured and computed masses, that have a typical size of
$\sim 0.5$ MeV, with the oscillatory contribution of the periodic orbits lying
in the chaotic components.

We are therefore led to evaluate the variance of the shell corrections that
originate in the chaotic layers. This can be done from the general
Eq.~(\ref{ut2f}). Again, our purpose is to make a statistical estimate valid
for a generic chaotic component, with no reference to a particular system. In
the range $\tau_{min} \ll \tau \ll \tauh$ the form factor of chaotic
single--particle levels is given by $K_D (\tau) = 2 \tau$ \cite{berry}. It
coincides with the random matrix prediction \cite{bgs}. This result is not
valid for times of the order of $\tau_{min}$. Extending nevertheless this
behavior down to $\tau \approx \tau_{min}$, and imposing again $K_D (\tau) =
0$ for $\tau < \tau_{min}$, we now arrive at
\begin{equation}\label{ut2c}
\langle \utc^2 \rangle = \frac{1}{8 \pi^4} \ E_c^2 \ . 
\end{equation}
Using Eq.~(\ref{ecn}), the typical size of the fluctuations $\sigma_{ch} =
\sqrt{\langle \utc^2 \rangle}$ of a chaotic component in the single--particle
dynamics is given by
\begin{equation} \label{sigc}
\sigma_{ch} = \frac{2.78}{A^{1/3}} \ {\rm MeV} .
\end{equation}
This expression has to be compared with the RMS of the difference $\delta U$
between the experimental and the computed masses, given in Ref.~\cite{mnms}.
The comparison is shown in Fig.~1.

The agreement is extremely good. The amplitude in Eq.~(\ref{sigc}) is
uncertain up to an overall factor of say, $2$. It can be varied by increasing
slightly the period of the shortest orbit (we have chosen the shortest
possible one; any modification will diminish $\sigma_{ch}$) and by the
inclusion of spin and isospin (this increases $\sigma_{ch}$ by a factor $2$ if
these components are treated as uncorrelated). The $A$ dependence is very well
fitted in the region $A \geqa 75$, with deviations observed for lower mass
numbers. This is in agreement with the limited accuracy of the Strutinsky
corrections and in general of semiclassical theories for light nuclei.

There are several features of the present theory that make its predictions
reliable. First of all, Eq.(\ref{ut2c}) contains only {\sl one physical
parameter}, the period of the shortest chaotic periodic orbit, which is a
function of $A$ because the size of the nucleus increases with the mass
number. But it has no dependence on the relative size of the chaotic region,
i.e., on the fraction of phase space occupied by chaotic motion. Without this
quite remarkable and important fact we would have been forced to estimate that
fraction, something that is hardly possible with present knowledge despite the
efforts in this direction since the pioneering work of Ref.\cite{arvieu}. This
is in contrast to what happens for the regular regions. The enhancement factor
$g$ present in Eq.~(\ref{ut2r}) -- but not in Eq.~(\ref{ut2c}) -- depends on
the corresponding {\it regular} average level density $\overline{\rho}_{reg}$.
The evaluation of $g$ in Eq.~(\ref{gn}) assumes that the nuclear
single--particle levels are regular, and the good qualitative agreement with
the experimental results of $\utr^2$ suggests that indeed most of the phase
space is occupied by regular trajectories. Mixed systems present therefore a
peculiar structure of shell effects. The amplitude of the fluctuations of the
regular phase space regions is proportional, through the factor $g$, to the
square root of the regular average level density, $\sigma_{reg} \propto
\sqrt{\overline{\rho}_{reg}}$. In contrast, the amplitude of the fluctuations
associated with the chaotic motion is constant, independent of their phase
space volume (aside from the transient nearly--integrable regime, not
considered here, where the chaotic layers are formed). In the extreme case of
a fully chaotic dynamics, $\overline{\rho}_{reg} = 0$ and $\ut = \utc$.

Equation (\ref{ut2c}) is in fact quite robust. It is not only independent of
the chaotic phase--space volume, but is also valid for arbitrary dimensions.
This fact suggests a very natural origin of these fluctuations, that we
haven't discussed yet. Although our analysis is based on a single--particle
picture, it can be extended to the full many--body phase space. In that space,
and in a semiclassical picture, to first approximation the point representing
the system follows a very simple trajectory driven by the regular mean--field
that dominates the motion. On top of that, it is likely that the residual
interactions, not taken into account in that approximation, induce chaotic
motion. The presence of chaotic orbits would then introduce additional
long--range modulations in the regular single--particle density of states.
Eq.~(\ref{ut2c}) would then still be valid to evaluate the amplitude of those
modulations, with $\tau_{min}$ the period of the shortest chaotic orbit in the
multidimensional space. Rough estimates indicate that $\tau_{min}$ is
comparable to the three--dimensional non--interacting period. Therefore
Eq.~(\ref{sigc}) presumably gives an estimate of the mass fluctuations arising
from neglected many--body effects.

To summarize, we have investigated shell effects of nuclear ground states
within a unified framework, namely periodic orbit theory. We show that a
dynamical symmetry--breaking mechanism that introduces chaotic layers in the
motion of the nucleons produces additional shell corrections to the total
energy. The intensity of the effect is, to first approximation, independent of
the size or the topology of the layers. It is governed by a single parameter,
$E_c$, which is proportional to the inverse time of flight of a nucleon across
the nucleus at Fermi energy. The comparison of the typical size of the chaotic
fluctuations is in very good agreement with the deviations between computed
and experimental values, for the amplitude as well as for their dependence
with the mass number. In contrast, the fluctuations produced by the regular
components are independent of the mass number, and are proportional to the
regular phase--space volume. The picture we are suggesting is coexistence of
order and chaos as produced, for instance, by residual interactions. It
manifests in nuclear low--energy properties. Further evidence should be given.
In particular, our theory predicts autocorrelations in the total energies
\cite{lm3}, as well as the effect of the presence of chaotic layers on the
total level density. This deserves further investigation.

\noindent We are indebted to W. J. Swiatecki whose questions and persistence
led to the present investigation, and to G. Bertsch and N. Pavloff for
enlightening discussions and comments. The Laboratoire de Physique Th\'eorique
et Mod\`eles Statistiques is an Unit\'e de recherche de l'Universit\'e Paris
XI associ\'ee au CNRS.

\begin{figure} \label{fig1}
\centerline{\psfig{figure=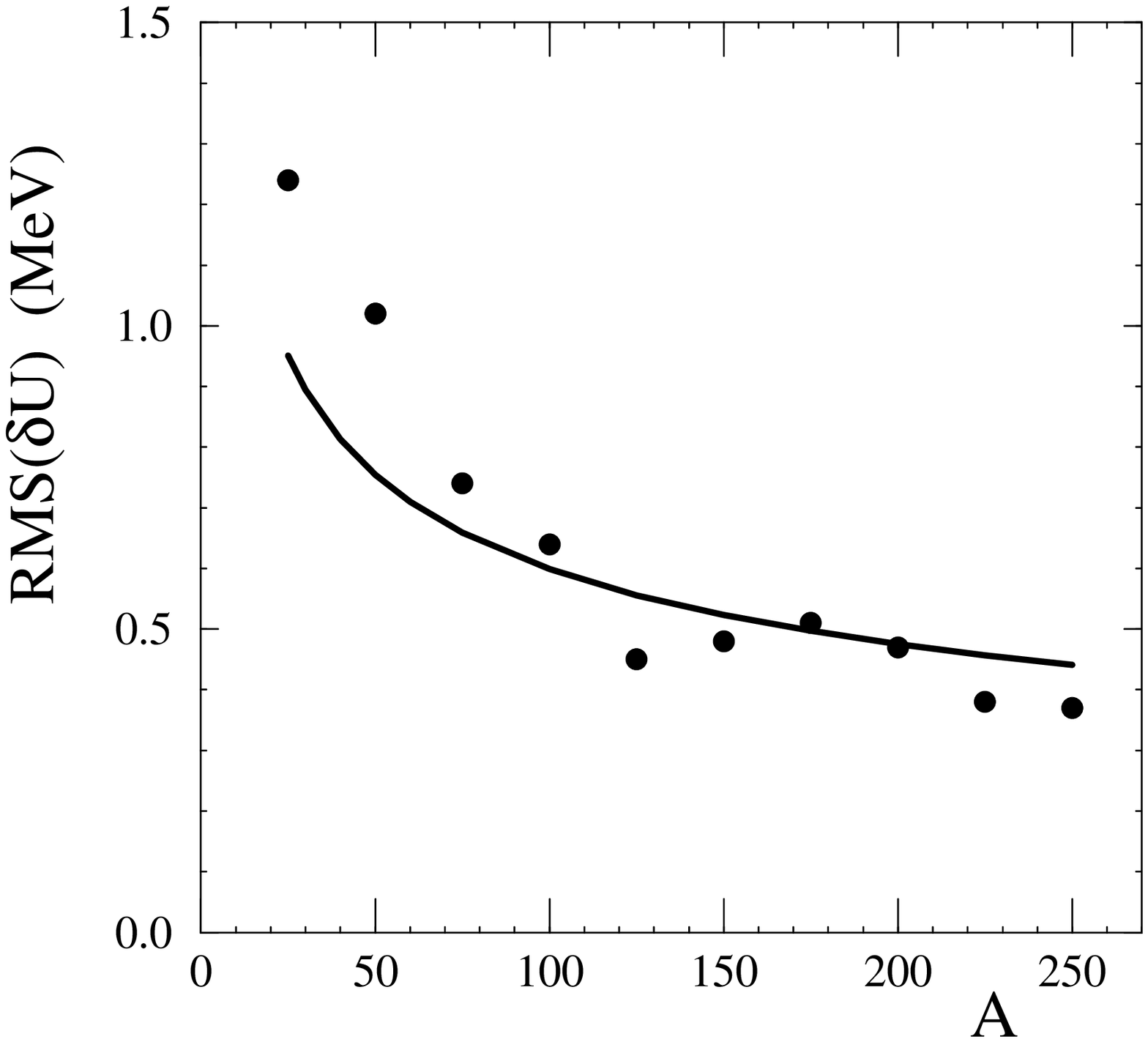,width=\figwidth,height=\figwidth}}
\caption{RMS of the difference $\delta U$ between computed and
observed masses as a function of mass number $A$. Dots taken from
Fig.~7 of Ref.~[3]. Solid curve from Eq.~(17).}
\end{figure}

\end{document}